\documentclass{PoS}
\usepackage{amsmath}

\title{Isospin breaking in 2+1 flavor QCD+QED}

\ShortTitle{Isospin breaking in 2+1 flavor QCD+QED}

\author{\speaker{Ran~Zhou}\thanks{We thank Riken, Brookhaven National
 Laboratory and the U.S. Department of Energy for providing the 
facilities and hospitality where this work was done.}\\
  Physics Department, University of Connecticut, Storrs, CT 06269-3046
 , USA\\
        E-mail: \email{zhouran@phys.uconn.edu}}

\author{Shunpei Uno\\
        Department of Physics, Nagoya University,
        Nagoya 464-8602, Japan\\
        E-mail: \email{uno@eken.phys.nagoya-u.ac.jp}}

\abstract{The mass splittings in the pseudoscalar mesons are
 studied by combining 2+1 flavor domain wall
fermion gauge configurations, generated by the RBC/UKQCD
collaborations, and quenched, non-compact,
lattice QED configurations. We extract the QED low energy
constants in SU(3) partially quenched chiral
perturbation theory, up to next-to-leading order, and
determine the non-degenerate u, d and s quark masses.
Systematic uncertainties due to chiral extrapolations are discussed.
The chiral symmetry breaking of domain wall fermions on the lattice
and its effect in partially quenched chiral
perturbation theory, including QED effects, is also
investigated.}

\FullConference{The XXVII International Symposium on Lattice Field Theory - LAT2009\\
		 July 26-31 2009\\
		 Peking University, Beijing, China}

\begin{document}
\section{Introduction}
The breaking of the strong isospin symmetry exists widely in hadronic 
physics. For example, the 
charged pion/kaon and neutral pion/kaon form 
a strong isospin multiplet. Their masses would be identical if 
strong isospin symmetry is preserved exactly. But we observe
$m^2_{\pi^+}-m^2_{\pi^0}\simeq m^2_{K^+}-m^2_{K^0}\neq 0$, which means 
this symmetry is broken in nature. The mass splitting in the hadron 
spectrum is due to different mass and charge of the fundamental 
quarks which compose the pion(kaon). The mass splitting in the hadron 
system can be used as a signature to detect the effect of QED 
interactions in QCD phenomena and has been studied by the lattice 
QCD community in recent years~\cite{Duncan:1996,Savage:2007,MILC}. 
It also provides a way to determine the non-degenerate u, d quark
masses, which can not be made from numerical simulations including QCD
interaction alone. The mass of the 
lightest quark, as a fundamental parameter of the Standard Model, 
is related to the 
strong CP problem and other important questions in high 
energy physics. 

The mass of the pseudoscalar meson is determined by both QCD and QED
dynamics. It is difficult to calculate the pseudoscalar mass spectrum
theoretically, because of the strong coupling in the QCD sector, which
makes perturbation theory invalid. Lattice QCD, on the 
other hand, is suitable tool to study hadronic physics 
non-perturbatively. We employ 2+1 flavor configurations generated 
by RBC/UKQCD collaborations~\cite{RBC:16.1,RBC:16.2,RBC:24} in this 
work, which include not only two degenerate light quarks, but also a heavy strange 
quark in the sea quark sector. Current lattice simulations can 
determine the pseudoscalar meson mass to roughly the 
1\% level(statistical error only). With the help of the 
average of $\pm e$ trick~\cite{Blum:2fEM}, we can determine 
 the mass-squared difference, 
$\delta m^2=m^2({\rm QCD+QED})-m^2({\rm QCD})$, to even 
higher accuracy. All of these enable us to build the connection 
between numerical simulation and theoretical description. For 
example, our result can be used to fit the Low Energy 
Constants (LEC's) in QCD+QED Partially Quenched Chiral Perturbation 
Theory (PQ$\chi$PT), which are difficult to obtain from experiment alone.

Preliminary results have been reported in~\cite{lattice:2008,TakuKAON09}. 
In this work, we present a new treatment of the violation of 
chiral symmetry and its further influence on the LEC's and quark mass.
We first give the PQ$\chi$PT description of the pseudoscalar meson 
mass in Sec.~\ref{sec:theory}. Explicit 
chiral symmetry breaking in the lattice regularization is discussed in 
Sec.~\ref{sec:chiral_symmetry}. The fit of QED LEC's and masses are 
shown in Sec.~\ref{sec:qmass}. The summary and plan for
future work is given in Sec.~\ref{summary}.\footnote{The mass
  splitting in the nucleon system will be reported 
separately in the future.}

\section{QCD+QED Partially Quenched Chiral Perturbation
  Theory  \label{sec:theory}}
Chiral Perturbation Theory($\chi$PT) is a low energy effective 
theory of QCD based on chiral symmetry. 
QCD+QED Partially Quenched Chiral Perturbation Theory(PQ$\chi$PT) 
gives the contribution of the sea quark and valance quark to the
pseudoscalar meson mass separately, which gives us more freedom 
to choose the mass of the sea and valance quarks in our lattice 
simulation. Suppose we have a kaon-like meson which is composed 
of u and s quarks. The mass and the charge 
of the quarks are $m_1$, $q_1$ and $m_3$, $q_3$ respectively. 
$q_i=2e/3,-e/3$ for up and down/strange quark.
So 
the square of the mass is~\cite{Bijnens:2007}:
\begin{eqnarray} 
m^2 =\chi_{13}+\frac{2C}{F_0^2}q_{13}^2+
\frac{\delta^{(4)}}{F^2_0}
\label{equ:m21}
\end{eqnarray} 
where $\chi_{13}= B_0(m_1+m_3)$, $q_{13}=(q_1-q_3)$, and
\begin{eqnarray} 
\frac{\delta^{(4)}}{F^2_0} &=&
 \left[(48 L_6^r - 24 L_4^r) \bar \chi_1 \chi_{13}
+  (16 L_8^r - 8 L_5^r) \chi_{13}^2 
-\frac{1}{3} \bar A(\chi_m) R^m_{n13} \chi_{13}
-\frac{1}{3} \bar A(\chi_p) R^p_{q\pi\eta} \chi_{13}\right]/{F^2_0}
\nonumber \\ &&
+   \bar A(\chi_{13}) q_{13}^2 +4   \bar
B(\chi_\gamma,\chi_{13},\chi_{13}) q_{13}^2
\chi_{13}-4  \bar B_1(\chi_\gamma,\chi_{13},\chi_{13}) q_{13}^2
\chi_{13}
\nonumber \\ &&
+{C}[-48 L_4^r q_{13}^2 \bar \chi_1
-16 L_5^r q_{13}^2 \chi_{13}
+2 \bar A(\chi_{1s}) q_{1s} q_{13}
-2 \bar A(\chi_{3s}) q_{3s} q_{13}]/F_0^4\nonumber \\ &&
- {Y_1} 4  \bar{Q_2}\chi_{13}+ {Y_2} 4 (q_1^2\chi_1+q_3^2\chi_3)+
{Y_3} 4  q_{13}^2 \chi_{13}
- {Y_4} 4   q_1 q_3\chi_{13}
+ {Y_5} 12  q_{13}^2 \bar \chi_1\nonumber \\ &&
\label{equ:m22}
\end{eqnarray}
The definition of symbols used in formula (\ref{equ:m22}) can be found 
in~\cite{Bijnens:2007}. $C$ is the ${\cal O}(e^2)$ low energy constant (LEC).
Our $q_i$ is  $e q_i$ of~\cite{Bijnens:2007}. Our ${\cal O}(e^2 m)$
LEC, $Y_i,(i=1,2,\cdots,5)$, are written in terms of linear
combinations of $K_j$ of~\cite{Bijnens:2007}.

\section{Violation of chiral symmetry on the lattice
  \label{sec:chiral_symmetry}}
Domain wall fermions (DWF) live in five-dimensional space and 
preserve exact chiral symmetry when the size of the extra 
dimension, $L_s$, goes to infinity. But the size of $L_s$ 
is always finite due to the limitation of the computational 
cost. For example, we set $L_s=$16 and 32 in our work, 
which means we do not have exact chiral symmetry in our 
simulation. The violation of chiral symmetry is quantified
by a small additive shift to the input quark mass, the residual mass, 
denoted as $m_{\rm res}$~\cite{RBC:24}. Thus the chiral limit is 
defined as $m_f+m_{\rm res}=0$, which leads us 
to replace all of the quark masses in 
the expression for the pseudoscalar mass in PQ$\chi$PT by $m_i+m_{\rm res}$~\cite{RBC:24}.

The analysis above is based on QCD interactions. If QED is included, 
it works as a perturbation to the residual mass which shifts 
$m_{\rm mres}$ by ${\cal O}(e^2)$ from its QCD value. 
Following usual perturbative renormalization arguments, 
the leading QED effect can be written as:
\begin{eqnarray}
\Delta m_{\rm{res}}\equiv m_{\rm{res}}({\rm QCD+QED})-m_{\rm
  {res}}({\rm QCD})=C_2(q_1^2+q_3^2),
\end{eqnarray}
where $q_1$ and $q_3$ are the charges of the quarks 
composing the pseudoscalar meson. The value of $C_2$ depends
on $L_s$ and vanishes exponentially as $L_s$ goes to infinity. Higher
order terms can be ignored at the order we are working, 
${\cal O}(e^2 m)$.

We next consider the effect of the residual mass to the pseudoscalar 
meson mass in the QCD+QED simulation. The QED contribution to $m_{\rm res}$ comes 
into the mass-squared of the pseudoscalar meson when combined with the leading
order QCD term, $2B_0C_2(q_1^2+q_3^2)$.
We have two different strategies to count this effect.
One is to measure the residual mass in the usual way on both QCD+QED 
and pure QCD configurations from the Ward-Takahashi Identity to get $\Delta m_{\rm{res}}$,
and extract $C_2$. Then we use $2B_0C_2(q_1^2+q_3^2)$ to account
for the QED induced violation of chiral symmetry.
 The other way is to add a term 
$\delta_{\rm{mres}} (q_1^2+q_3^2)$ to 
the expression of $m^2$ of the pseudoscalar meson, which means we 
leave $2B_0C_2$ as new free parameter in PQ$\chi$PT formula. We 
can compare whether $\delta_{\rm mres}$ is consistent with $2B_0 C_2$, which is a way to check whether our understanding of the
QED violation of chiral symmetry is correct or not.

\begin{figure}
  \centering \includegraphics[scale=0.75]{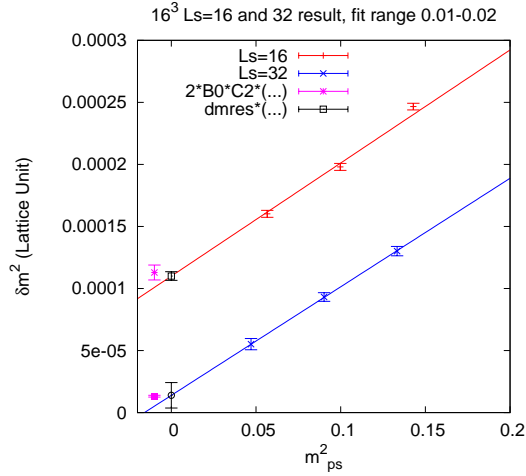}
\caption{$\delta m^2$ of $d\bar d$ meson with Ls=16 and 32. dmres(...)
  means $\delta_{\rm mres}(q_1^2+q_3^2)$. It's value at $m_{ps}=0$
  comes from extrapolation. 2B0C2(...) means $2B_0 C_2(q_1^2+q_3^2)$.
  (It is shifted horizontally towards left to make plot clear.) 
  $C_2$ is determined from the change of the residual mass by 
  setting $e\neq 0$ and $e=0$. The error bar of this term 
  comes mostly from the error in $B_0$.}
\label{fig:ls32mres}
\end{figure}

We test our understanding about the violation of the chiral symmetry
on $16^3$ lattice configurations with $L_s$=16 and 32. The result 
is shown in figure~\ref{fig:ls32mres}. The upper line corresponds to 
the $L_s$=16 case; we measure $\delta m^2=m^2({\rm QCD+QED})-m^2({\rm
  QCD})$ of the $d\bar d$ meson with $m_{val}=$0.01, 0.02 and 0.03 
and physical charge and extrapolate to $m_f+m_{\rm res}{\rm (QCD)}=0$ 
to get $\delta_{\rm{mres}}(q_1^2+q_3^2)$. We also determine $C_2$ 
by measuring the change in $m_{\rm res}$ when 
$e\neq 0$ and $e=0$. Then the value of $2B_0 C_2(q_1^2+q_3^2)$ 
is calculated and shown on the plot (shifted left 
a little to make the plot clear). The error of $2B_0 C_2(q_1^2+q_3^2)$ 
term mostly comes from the error on $B_0$. It is clear from 
figure~\ref{fig:ls32mres} that these two results are
consistent with each other. This consistency is also checked for the $L_s=32$ case. Because the violation of the 
chiral symmetry of DWF is decreased when 
$L_s$ goes to larger value, it is expected that the 
contribution of the $\delta_{\rm mres}(q_1^2+q_3^2)$
and $2 B_0 C_2(q_1^2+q_3^2)$ should be suppressed at 
larger $L_s$. It 
can be seen from the plot that both the $\delta_{\rm mres}$ and $C_2$ terms are decreased by about an order of magnitude 
at the larger value of $L_s$, but their contributions are still 
consistent within the error bar. All of these confirm the
validity of using $\delta_{\rm mres}(q_1^2+q_3^2)$ to account 
for the contribution of the violation of the chiral symmetry.

\section{Numerical Result on Low Energy Constant and Quark Mass
\label{sec:qmass}}
 Formula (\ref{equ:m21}) can be used to write the mass-squared splitting as:
\begin{eqnarray}
\delta m^2 & = \delta m^2(\rm{physical}) + \delta m^2(\rm{lattice\ \
  artifact}) 
\end{eqnarray}
where:
\begin{eqnarray}
\delta m^2(\rm{physical}) &=&  {2C}q_{13}^2/{F_0^2}
+   \bar A(\chi_{13}) q_{13}^2 +
4 \bar B(\chi_\gamma,\chi_{13},\chi_{13}) q_{13}^2
\chi_{13}-4  \bar B_1(\chi_\gamma,\chi_{13},\chi_{13}) q_{13}^2
\chi_{13} 
\nonumber \\ &&
 +{C}[-48 L_4^r q_{13}^2 \bar \chi_1
-16 L_5^r q_{13}^2 \chi_{13}
+2 \bar A(\chi_{1s}) q_{1s} q_{13}
-2 \bar A(\chi_{3s}) q_{3s} q_{13}]/F_0^4\nonumber \\ &&
- {Y_1} 4  \bar{Q_2}\chi_{13}+ {Y_2} 4  q_p^2\chi_p+
{Y_3} 4  q_{13}^2 \chi_{13}
- {Y_4} 4   q_1 q_3\chi_{13}
+ {Y_5} 12  q_{13}^2 \bar \chi_1 
\end{eqnarray}
and 
$\delta m^2(\rm{lattice\ \ artifact})  = \delta_{\rm{mres}}
(q_1^2+q_3^2)$

\begin{figure}
  \includegraphics[scale=0.6]{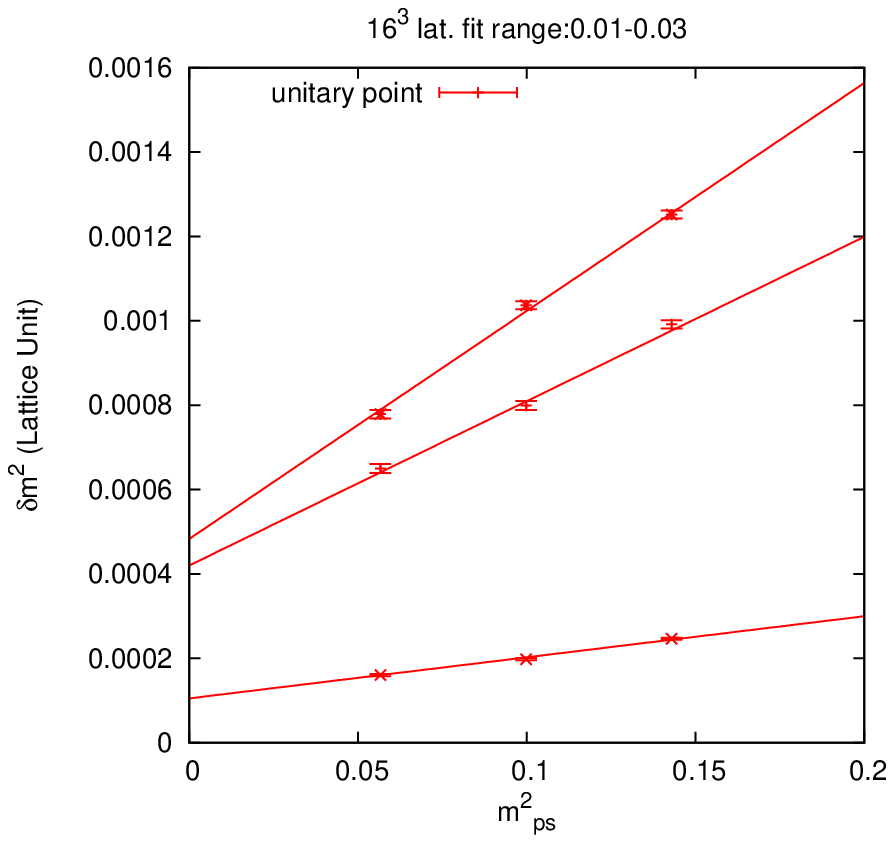}
  \includegraphics[scale=0.6]{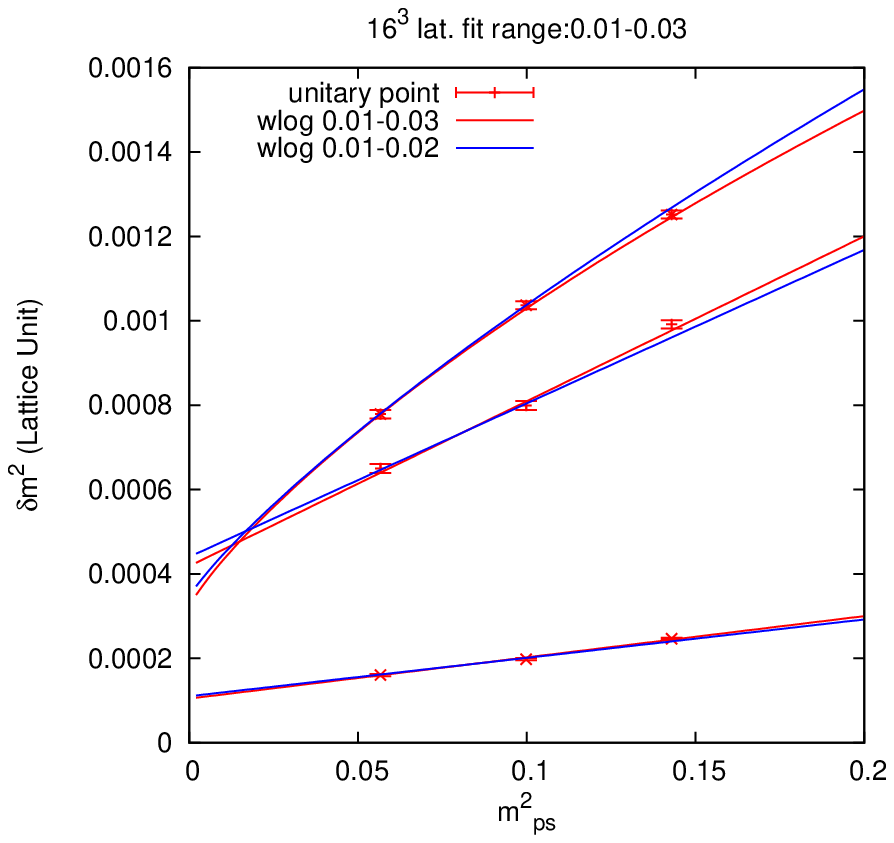}
  \includegraphics[scale=0.6]{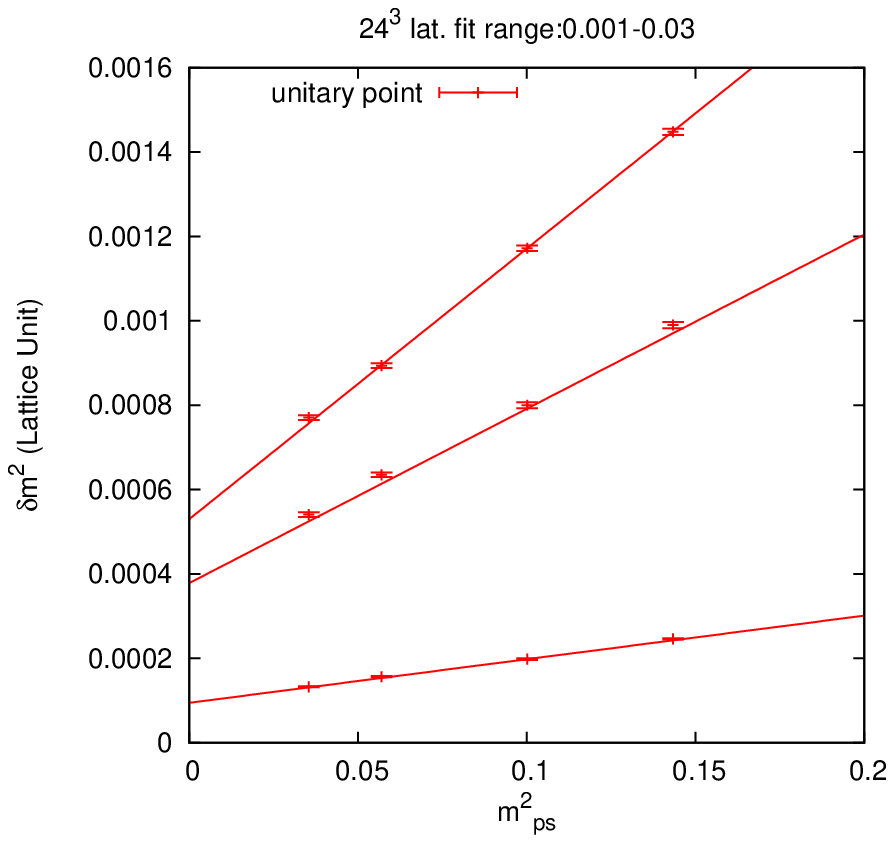}
  \includegraphics[scale=0.6]{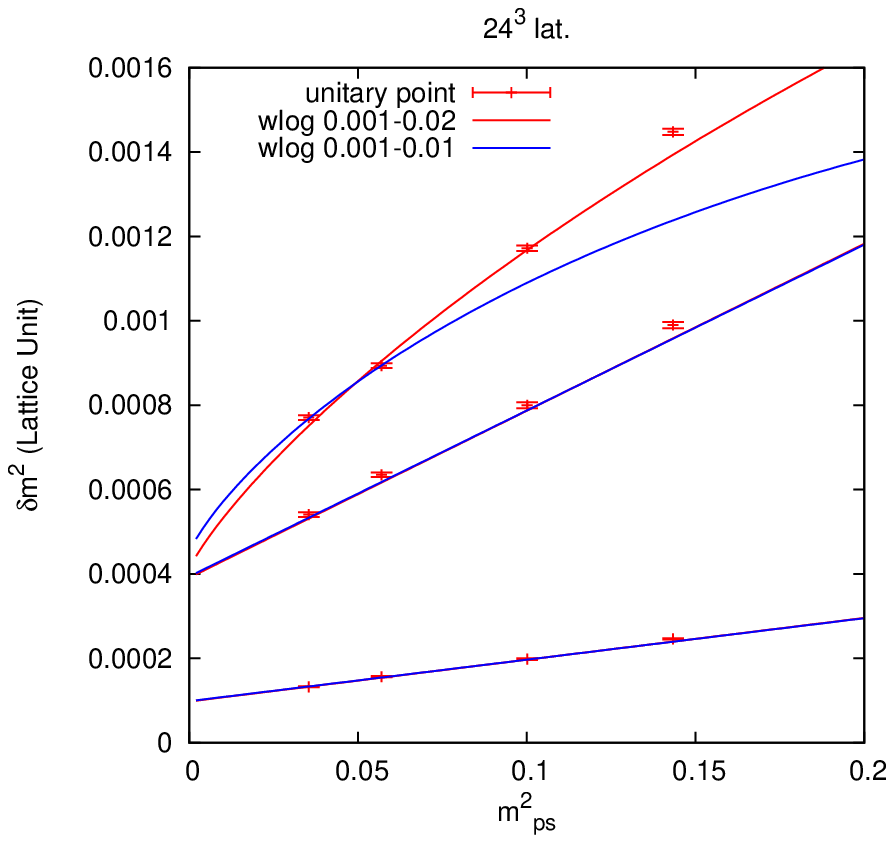}
  \caption{$16^3$ linear fit(upper left panel), $16^3$ chiral 
    log fit(upper right panel), $24^3$ linear fit(lower left panel) 
    and $24^3$ chiral log fit(lower right panel) lattice data points 
    and fit results for the meson mass splitting. Lines correspond 
    to fits with and without chiral logs. 
    The data points in the plot correspond to $u\bar d, u\bar u$ and 
    $d\bar d$ mesons, respectively, from top to bottom.}
\label{Fig:no_sub.fit}
\end{figure}

By fitting the results of $\delta m^2$, we can obtain all of the QED
LEC's, including $\delta_{\rm{mres}}$. The fit results are shown in 
figure~\ref{Fig:no_sub.fit} along with the 
unitary data points (though all of the partially quenched data points
within mass cut were used in the fit).
From this figure, we can see that charged meson receives more
finite volume corrections than the neutral meson, which is consistent
with the theoretical investigation~\cite{Hayakawa:2008an}.
We perform fits with and without the log terms to probe the effect 
of the chiral logarithms. $\chi^2/dof$ is adequate in either case, except for
$24^3$ when we include quark masses larger than 0.02 in the
fit. One of the reason is that the $\chi^2/dof$ for the 
pseudoscalar meson mass fit is around 2 at heavier quark mass, 
which is a little bit large. The other reason is that the quark 
mass $m_{q}=$0.02 or 0.03 may be too heavy for SU(3) 
PQ$\chi$PT~\cite{RBC:24}. We need to choose the fit range 
carefully. If the heavier sea quark points are omitted, $\chi^2$ becomes 
reasonable. The main effect of the logs is to significantly 
reduce the value of the charged meson splitting in the chiral 
limit (so-called Dashen term). Note, the log terms vanish for 
the neutral meson splittings at NLO. We put the best result of 
QED LEC's as we have in Table~\ref{tab:LEC} and compare with
the phenomenological value~\cite{Bijnens:2007}. Our LEC's come from 
$24^3$ lattices fit , which receive less finite volume correction. 
The fit range is 0.001-0.01\footnote{We add $m_{\rm val}=0.001$ data 
to our analysis. So our fit result are changed form those reported 
in~\cite{lattice:2008} as well the result quoted in~\cite{TakuKAON09}, 
where only 0.005 and heavier quark masses are used.}, which make 
SU(3) PQ$\chi$PT to be valid.
The log terms are also included in the fit.

\begin{table}[ht]
 \centering
 \begin{tabular}{|c|c|c|c|c|c|c|c|}
  \hline  
  & $10^6C$ & $10^2Y_2$ & $10^3Y_3$ & $10^3Y_4$ & $10^2Y_5$ &
  $10^3\delta_{\rm mres}$ & $\chi^2/dof$ \\ \hline 
  $24^3$ fit & 0.27(19) & 1.59(10) & -10.6(7) & 9.8(16) & 2.00(68) 
  & 5.08(9) & 2.11(73)  \\ 
  Ref.~\cite{Bijnens:2007} & 7.3 & 0.38 & 1.58 & 2.83 & -0.953 & N/A 
  & N/A \\ \hline
 \end{tabular}
 \caption{The preliminary result of the QED LEC's from fits of 
   $\delta m^2$ and only statistical error is quoted. The QCD LEC's 
   were taken from the SU(3) PQ$\chi$PT fit by RBC/UKQCD collaborations. All
   of the QCD and QED LEC's are defined at scale $\mu=1$GeV. 
   The last row gives phenomenological estimate
   in~\cite{Bijnens:2007}.}
 \label{tab:LEC}
\end{table}

Next, we determine the physical u,d, and s quark masses from the QCD LEC's
given by the RBC/UKQCD collaborations and the QED LEC's from our fit. 
The fluctuation in QCD configurations is considered by including the jack
knife blocks of QCD LEC's. Since there are three unknown 
quark masses, we need the experimental value of the mass from 
three pseudoscalar mesons ($\pi^+$, $K^+$ and $K^0$) to fix the quark 
masses~\cite{Amsler:2008zzb}. The mass of the 
neutral pion is not used, because we don't include the disconnected 
diagram in our simulation due to the difficulty of computation. 
The renormalized quark masses are defined as 
$m_q^{\overline{\rm MS}}=Z_m(m_q+m_{\rm res})$, where $Z_m$ is 
the matching factor from lattice to $\overline{\rm MS}$(NDR) scheme
at a scale $\mu=2$ GeV obtained via  the  RI/MOM scheme with the
non-perturbative technique\footnote{Here our $Z_m$ is calculated
on pure QCD configurations.}. We use the value $Z_m=1.656(48)(150)$ 
from~\cite{RBC:24}.
Table~\ref{tab:qedmass} gives the result of the quark masses. 
The analysis on the $24^3$ lattice has less finite volume 
effect and reasonable $\chi^2$  for the  LEC's fit for the fit range
0.001-0.01. So we quote the quark mass determined from these
parameters as our preliminary result, 
$m^{\overline{\rm MS}}_u=2.79(37)$MeV, 
$m^{\overline{\rm MS}}_d=4.84(52)$MeV and 
$m^{\overline{\rm MS}}_s=95.9(9.6)$MeV.

\begin{table}[ht]
 \centering
 \begin{tabular}{|c|c|c|c|c|c|c|}
  \hline  
  lattice & fit range & $m_u$ & $m_d$ & $m_s$ &
  $\frac{m_u}{m_d}$ & $\frac{m_s}{m_{ud}}$\\
  \hline
  $24^3$ & 0.001-0.01 & 2.79(37) & 4.84(52) & 95.9(9.6) & 0.57(1) &
  25.1(5) \\
  \hline
 \end{tabular}
 \caption[]{Preliminary results of the u, d and s quark masses
   determined from pure QCD interaction. 
   The value is in the unit of MeV and $\overline{\rm MS}$(NDR) scheme 
   at renormalization scale $\mu=2$ GeV. The QCD LEC's were taken 
   from the SU(3) fit by RBC/UKQCD collaborations and QED LEC's are
   from Table~\ref{tab:LEC}. The error quoted here is only statistical 
   error.}
 \label{tab:qedmass}
\end{table}

\section{Summary and Plan of Future Work\label{summary}}
Our simulations are based on 2+1 flavor QCD configurations 
generated by RBC/UKQCD collaborations and quenched 
non-compact QED configurations generated by us.  After 
treating the violation of chiral symmetry carefully, we fit 
the pseudoscalar meson mass-squared differences using QCD+QED PQ$\chi$PT to 
extract the QED LEC's. The LEC's are affected by 
the volume of the lattice, fit range and the formula used 
in the fit. We have investigated these systematics, but have not yet 
quoted errors for them in our preliminary results. 
The up, down, and strange quark masses are also determined
using the LEC with the experimental mass values for pseudoscalar
mesons (pi+, K+, K0). Our preliminary results for $\overline{\rm MS}$(NDR)
scheme at the renormalization scale $\mu=2$ GeV is obtained using
LEC's determined on 24$^3$ lattice and fit range as 0.001-0.01. They are
$m^{\overline{\rm MS}}_u=2.79(37)$MeV, 
$m^{\overline{\rm MS}}_d=4.84(52)$MeV and 
$m^{\overline{\rm MS}}_s=95.9(9.6)$MeV. Errors are statistical only. 

In this work, quenched QED configurations were used to 
account for the EM interaction. The systematic error caused 
by this approximation can be removed by the reweighing
method~\cite{Duncan:2004ys}. In addition, the analysis on different lattice
spacing will allow to extrapolate to the continuum
limit. This will be done on the companion 32$^3$ lattice ensemble generated by the RBC and UKQCD collaborations (see the talks by Kelly and Mawhinney in these proceedings).

\section{Acknowledgments}
This paper is based on the collaborations with T. Blum, T. Doi, 
M. Hayakawa, T. Izubuchi and N. Yamada.
We thank the US Department of Energy and RIKEN for the support
necessary to carry out this research. RZ  and TB were supported by US
DOE grant  DE-FG02-92ER40716. NY is supported by the Grant-in-Aid of
the Ministry of Education(No. 20105001, 20105002). TD is supported  in
part by Grant-in-Aid for JSPS Fellows 21$\cdot$5985. Computations 
were carried out on the QCDOC supercomputers at the RIKEN
BNL Research Center, BNL, and Columbia University.


\begin{thebibliography}{99}


\bibitem{Duncan:1996}
 A.~Duncan, E.~Eichten and H.~Thacker, 
 \emph{Electromagnetic splittings and light quark masses in lattice
   QCD}, 
 Phys.\ Rev.\ Lett.\  {\bf 76}, (1996) 3894, 
 arXiv:hep-lat/9602005.

\bibitem{Savage:2007}
  S.~Beane, K.~Orginos and M.~Savage
 \emph{Strong-isospin violation in the neutron proton mass
   difference from  fully-dynamical lattice QCD and PQQCD}
 Nucl.\ Phy.\ B.\  {\bf 768}, (2007) 38-50,
 arXiv:hep-lat/0605014.

\bibitem{MILC}
S. Basak {\it et al.,}
  \emph{Electromagnetic splittings of hadrons from improved staggered quarks
  in full QCD}
  arXiv:0812.4486 [hep-lat] \\


\bibitem{RBC:16.1}
  D.J.~Antonio {\it et. al}
  \emph{First results from 2 + 1 - flavor domain wall QCD: Mass
    spectrum,
 topology change, and chiral symmetry with $L_S = 8$}
 Phys.\ Rev.\ D.\  {\bf 75}, (2007) 114501, 
 arXiv:hep-lat/0612005.

\bibitem{RBC:16.2}
  C.~Allton {\it et. al}
 \emph{2 + 1 flavor domain wall QCD at $(2fm)^3$ lattice: Light meson
   spectroscopy with $L_S =  16$}
 Phys.\ Rev.\ D.\  {\bf 76}, (2007) 014504,
 arXiv:hep-lat/0701013.

\bibitem{RBC:24}
C.~Allton {\it et. al}
 \emph{Physical Results from 2+1 Flavor DomainWall QCD and SU(2)
   Chiral Perturbation Theory}
 arXiv:hep-lat/0804.0703.

\bibitem{Blum:2fEM}
  T.~Blum, T.~Doi, M.~Hayakawa, T.~Izubuchi, and N.~Yamada
 \emph{Determination of light quark masses from the electromagnetic
   splitting of pseudoscalar meson masses computed with two flavors 
   of domain wall fermions}
 Phys.\ Rev.\ D.\  {\bf 76}, (2007) 114508,
 arXiv:hep-lat/0708.0484.

\bibitem{lattice:2008}
  R. Zhou, T. Blum, T. Doi , M. Hayakawa , T. Izubuchi, N. Yamada
 \emph{Isospin symmetry breaking effects in the pion and nucleon masses}
 arXiv:0810.1302 [hep-lat]

\bibitem{TakuKAON09}
  T.Izubuchi 
  \emph{Studies of the QCD and QED effects on Isospin breaking}
       {\bf PoS, KAON09} 2009,034

\bibitem{Bijnens:2007}
  J.~Bijnens and N.~Danielsson
 \emph{Electromagnetic corrections in partially quenched chiral
   perturbation theory}
 Phys.\ Rev.\ D.\  {\bf 75}, (2007) 014505, 
 arXiv:hep-lat/0610127.


\bibitem{Hayakawa:2008an}
  M.~Hayakawa and S.~Uno,
  \emph{QED in finite volume and finite size scaling effect on
    electromagnetic
  properties of hadrons}
  arXiv:hep-ph/0804.2044.


\bibitem{Duncan:2004ys}
  A.~Duncan, E.~Eichten and R.~Sedgewick,
  Phys.\ Rev.\  D {\bf 71}, 094509 (2005)
  [arXiv:hep-lat/0405014].



\bibitem{Amsler:2008zzb}
Amsler {\it et al.}
  \emph{Particle Data Group}
  Phy.\ Lett.\ B {\bf 667}, 1 (2008)
\end{thebibliography}
\end{document}